# Smart Analytical Signature Verification For DSP Applications

Rozita Teymourzadeh, *CEng, Member IEEE/IET*, Waidhuba Martin Kizito, Kok Wai Chan, and Mok Vee Hoong
Faculty of Engineering, Technology & Built Environment
UCSI University
Kuala Lumpur, Malaysia
rozita@ucsiuniversity.edu.my

*Abstract* — Signature verification is an authentication technique that considers handwritten signature as a "biometric". From a biometric perspective, this project made use of automatic means through an integration of intelligent algorithms to perform signal enhancement function such as filtering and smoothing for optimization in conventional biometric systems. A handwritten signature is a 1-D Daubechies wavelet signal (db4) that utilizes Discrete Wavelet Transform (DWT) and Discrete Cosine Transform (DCT) collectively to create a feature dataset with d-dimensional space. In the proposed work, the statistical features characteristics are extracted from each particular signature per data source. Two databases called Signature Verification Competition (SVC) 2004 database and SUBCORPUS-100 MCYT Bimodal database are used to cooperate with the design algorithm. Furthermore, dimension reduction technique is applied to the large feature vectors. A system model is trained and evaluated using the support vector machine (SVM) classifier algorithm. Hence, an equal error rate (EER) of 8.7% and an average correct verification rate of 91.3% are obtained.

## I. INTRODUCTION

Signature forgery is a major concern for financial institutions around the world. About 83.6% of the financial institution respondents in Dallas Fed District reported experiencing the highest number of fraud attempts for payments involving signature debit cards [1]. Furthermore, around 86% of the financial institution respondents have identified payments using signature debit cards as having the highest dollar losses due to fraud [1]. Since signatures are commonly required for validation of cheques, debit cards, credit cards and historical financial documents, there is a need to develop a robust signature verification system to identify the owner of the signature and to determine if the signature is genuine. Hence, this research work proposes an automatic online signature verification system based on both Discrete Cosine Transform (DCT) and Discrete Wavelet Transform (DWT) to be implemented for digital signal processing (DSP) applications.

Signature verification system can be classified into two categories: off-line and on-line signature verification [2, 3]. In recent years, the growing interest towards personal identity authentication along with increase demand for security issues has led to the development of various signature verification systems [4-8]. The performance of these signature verification systems are often measured by its false rejection rate of genuine signatures (FRR), false acceptance rate of forgery signatures (FAR), and the equal error rate (EER), which is the point where the false acceptance rate and the false rejection rate are the same.

In 2008, Nanni and Lumini [4] proposed the use of DWT along with DCT for online signature verification to extract signatures' local information. The result of this system, based on the MCYT-100 database, is an EER of 11.5% [4]. Kholmatov and Yanikoglu [5], in July 2009, presented an online signature verification using global features consisting of Fourier Descriptors, which yields an improved result of 7.22% EER for the MCYT-100 database. More recently, Wang et al. [6] proposed using wavelet packet (WP) for online signature verification system. The signature verification system, when tested on the SVC2004 database, yields an average EER of 6.65% in skilled forgery detection [6] and an average EER of 1.44% in random forgery detection [6]. His research work shows that, by applying both DWT and DCT to the signature verification which creates a feature dataset with d-dimensional space, along with further dimension reduction by Principal Component Analysis (PCA), the signature verification will perform better in its EER and average correct rate. The detailed fundamental stage realization of signature verification will be discussed and scrutinized.

The paper is organized as follows. Section II and III present the respective fundamentals the proposed signature verification system, along with the system implementation. Experiments conducted on the SVC2004 and MCYT-100 databases of online signatures are presented and discussed in Section IV. Section V reports the conclusion of the paper.

## II. SYSTEM REALIZATION

In this chapter, the overall system structure is given. Fig. 1 shows the fundamental of signature verification system. In the signal processing stage of the biometric system, a set of algorithms is utilized to locate and extract the desired biometric features from data, while ensuring fast processing and accurate results. However, the modern technology approaches such as Fast Fourier Transform (FFT), DCT and DWT are applied to coordinate for digitalization and evaluate the raw digital data. The signature is digitalized and presented as signal time sequence.

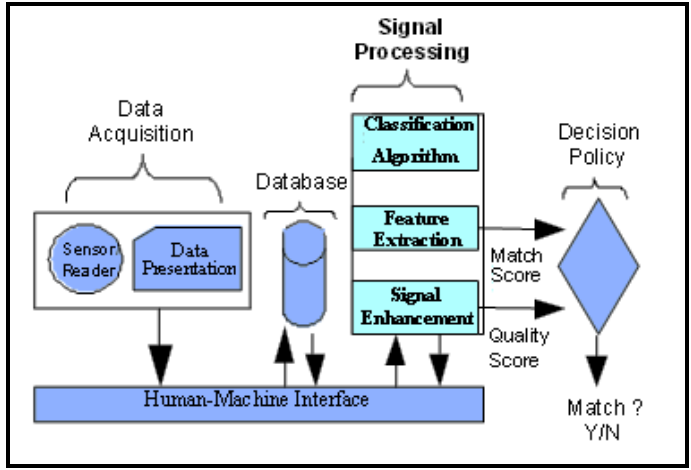

Fig. 1. Overview of a biometric system.

Along with digital processing, the signature is sampled and stored in database and (1) shows $n$–sample of signature in digital format [5].

$$S(n) = [x(n)y(n)t(n)p(n)]^T \qquad (1)$$

where $x(n)$ and $y(n)$ denote the coordinates of the points on the signature trajectory, while both $t(n)$ and $p(n)$ indicate the pen pressure and timestamp, at sample point $n$. Data collected from well-organized signature databases are used to lay focus on the signal processing task of the system.

Wavelet analysis is preferred for the signal preprocessing because, unlike in time domain or frequency domain specific applications, both the time localization and frequency localization of a signal can be obtained. Wavelets cut up data into frequency components before analyzing each frequency component with a resolution matched to its scale. These wavelets are defined in reference to a "mother function" $\psi$, as demonstrated in (2) [7]. By translating and scaling the mother wavelet, a whole family of wavelets can be generated [7].

$$\psi_{(a,b)}(t) = \frac{1}{\sqrt{a}} \psi\left(\frac{t-b}{a}\right), \; a>0, b \in \mathbb{A} \qquad (2)$$

In (2) and (3), $a$ represents the scaling parameter, whereas $b$ denotes the translation parameter. The coefficients of (3) measure the variations of field $f(t)$ about the point $b$, with the scale represented by $a$. A pyramid algorithm is used to conduct the wavelet analysis on the discrete data.

$$\psi_{(a,b)} = \int_{-\infty}^{\infty} f(t)\psi_{(a,b)}(t)dt \qquad (3)$$

DWT is found to yield faster computation of the wavelet transform. The implementation of the DWT is easy, with reduced computation time. The signal can be transformed to time and harmonic domain using DWT and filtering in discrete time domain. Later, the signal will be analyzed with different cutoff frequencies at different scales.

A sequence of data points can also be reconstructed very accurately from only a few DCT coefficients, which is useful for applications that require feature data reduction in this research work.

As the data do not lie perfectly on a linear subset, some of the information contained in the original data will be lost during data compression. Hence, in order to find the compression directions that will result in the least amount of information is lost from the original data, the method of Principal Component Analysis is applied. In Fig. 2, PCA identifies the two directions (PC1 and PC2) along which the data have the largest spread.

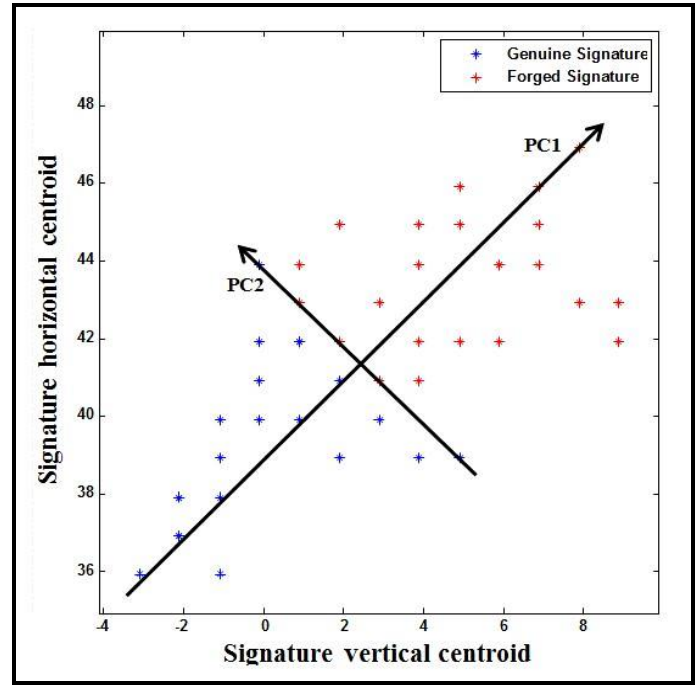

Fig. 2. Principal Component Analysis (PCA) of a signature data set.

PCA allows the computation of a linear transformation that maps data from a high dimensional space to a lower dimensional space, as shown in (4).

$$b_k = t_{k1}a_1 + t_{k2}a_2 + ... + t_{kN}a_N \qquad (4)$$

where $a$ refers to the feature dataset in higher dimension space, whereas $b$ refers to the feature dataset in lower dimension space, and $t$ denotes the projection constant at feature position $k$.

The objectives of our research work have been narrowed to design and develop a smart analytical tool for signature verification and identification. The design incorporates both the wavelet toolbox from Signal Processing block set and the Bioinformatics block set to identify authentic digital signature.

### III. IMPLEMENTATION

From Fig. 3, load signature dataset is first selected from one of the data sources for processing, either from the MCYT-

100 database, SVC2004 database or signature tablet. During the pre-processing stage, normalization is performed with the preferred algorithm to attain a single feature vector to be fed into the support vector machine (SVM) classifier.

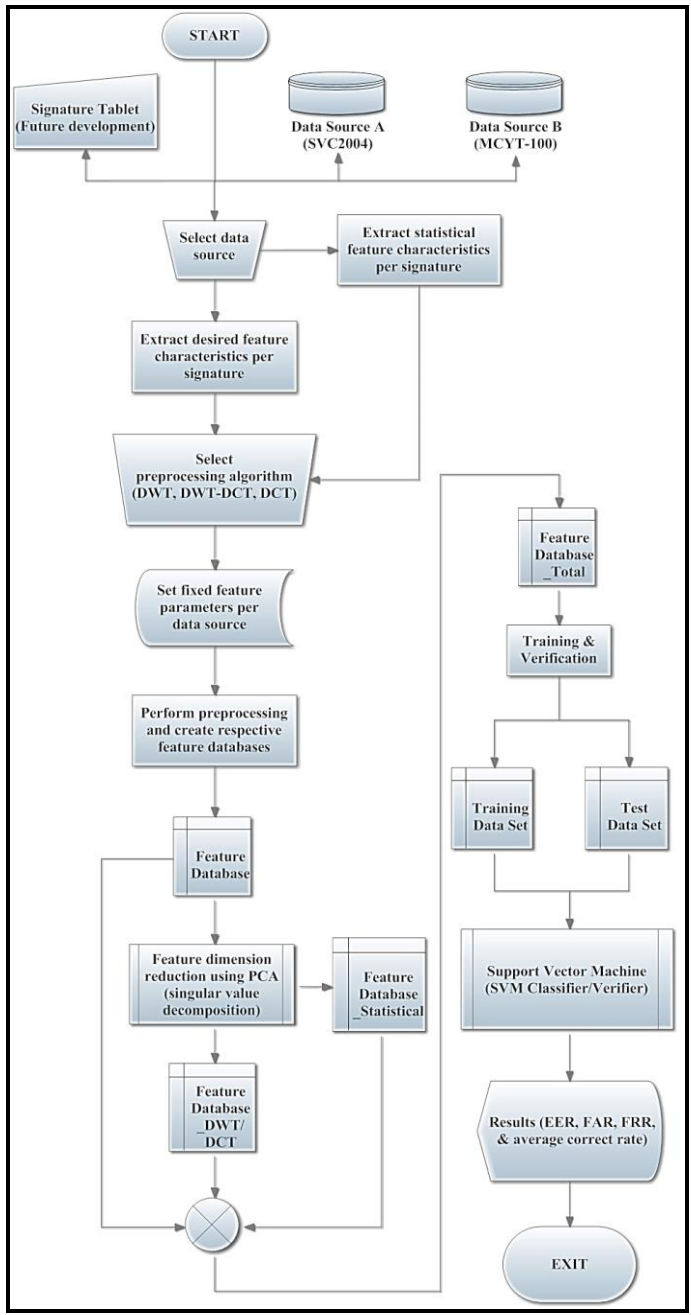

Fig. 3. Signature verification algorithm flowchart.

The pre-processing algorithms available are DWT, DCT, and the combined DWT-DCT. Filtering and smoothing are applied for enhancing the signal. With feature extraction, the raw data is transformed into a processable dataset with d-dimensional feature space. Since the statistical features, which are extracted on each signature sample piece, are acquired, the feature vector becomes so large that it requires some dimension reduction which is performed using the PCA technique [20]. A model is then trained using the available data and its performance is evaluated using outputs of the SVM classifier algorithm and the real state of the signature, whether it is a genuine signature or a forgery signature. A test-run was performed and the proposed system was evaluated.

Signal enhancement using wavelet toolbox for wavelet preprocessing level discrete computes a single level discrete 1-D wavelet transform with respect to a particular wavelet decomposition filter. In our research work, the Daubechies wavelet series have been utilized to implement the decomposition stages. The wavelet analysis performed on a single x-coordinate value extracted from the signature dataset is demonstrated in Fig. 4.

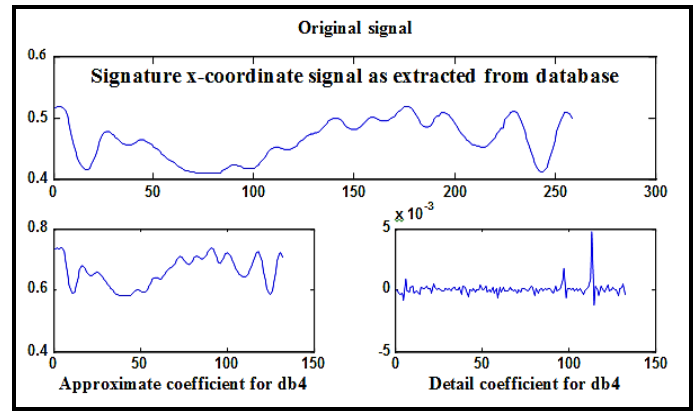

Fig. 4. A wavelet signal extracted using x-coordinate value of a signature sample (MCYT-100 signature database).

In the analysis of DWT, filter is applied to a single signal extracted from the signatures feature set, either x-coordinate or y-coordinate. Observation of output signifies that the Daubechies wavelet of order 4 (dB4) performs relatively better than other orders of Daubechies wavelet. Fig. 5 depicts the example of results attained from the SVC2004 sample signature database. The decomposition low-pass (Lo_D) filter and decomposition high-pass (Hi_D) filter are the key focus of our research work with respect to the application of signature verification. It is required that the signal be decomposed and passed to the data classifier for feature training.

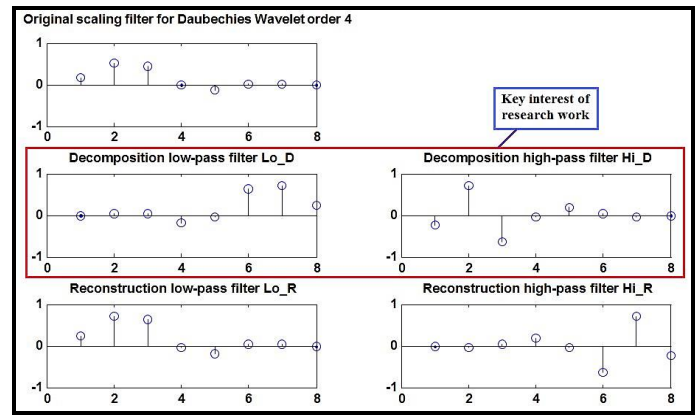

Fig. 5. Daubechies wavelet filter decomposition (db4).

As portrayed in Fig. 6, the designed user interface for the proposed algorithm comprises of the signature data acquisition, data preprocessing, and classification and verification. The system user will first select a data source, a signer, and the corresponding signature sample category (either forgery or genuine) to access the raw data. A preferred data-preprocessing algorithm is then selected to initialize the system operation. Both the signature feature image(s) and verification status analysis are displayed in the system interface.

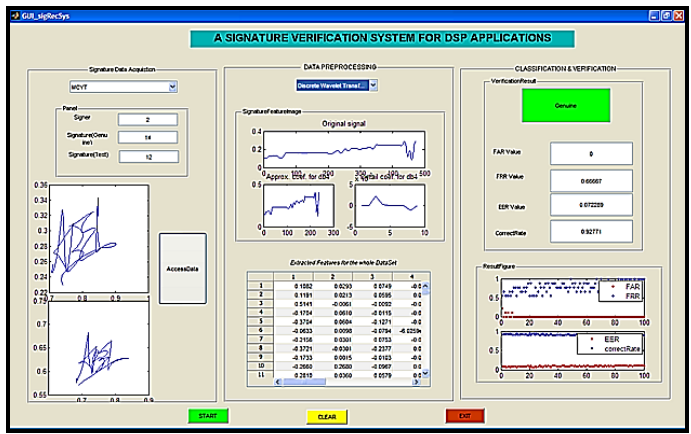

Fig. 6. Test run on the proposed signature verification system.

## IV. RESULTS AND DISCUSSIONS

In this research work, the two signature databases of Subcorpus-100 MCYT and SVC2004 were used for the training and testing purposes.

With Subcorpus-100 MCYT signature database support, PCA technique was employed to locate the most useful feature points for the classification stage. Few parameters were taken into consideration in the proposed research, that are x-coordinate, y-coordinate, pen altitude and pen elevation sequence of the pen with respect to the tablet. After the DWT transformation, 53 values are accessed for each of the nine parameters. As a result, there are 477 feature points available to be used for classification purpose. With the application of the PCA technique, the feature points are reduced to the 8 top feature components and retained. The PCA technique is also applied to the statistical features, retaining the top 2 features. In both cases, the eigenvalue contribution is 1.00. SVM classifier is then used to realize the final classification.

The following information displayed on Table 1 was extracted from the proposed system after the signature was injected to the algorithm.

Table 1. Sample signature verification results from the MCYT-100 signature database for Signer A and Signer B.

| Signer | EER (%) | FAR (%) | FRR (%) |
|---|---|---|---|
| Signer A | 8.43 | 22.22 | 55.56 |
| Signer B | 7.23 | 0 | 66.67 |

As shown in Table 1, the FAR may relate to the system not having full accurate tools, and the FRR may refer to the user error. Hence, the FAR restriction is justified in the proposed algorithm.

The SVC2004 database was also applied to verify the proposed signature verification system. Compared to the MCYT-100 database, the SVC2004 database provides reliable data features. However, both pressure from signing and timing factor were analyzed in proposed research work. The same steps of model training and testing on the MCYT-100 database are also performed on the SVC2004 database. The results attained are presented in Table 2.

Table 2. Sample signature verification results from the SVC2004 signature database for Signer A and Signer B.

| Signer | EER (%) | FAR (%) | FRR (%) |
|---|---|---|---|
| Signer A | 8.70 | 0 | 57.14 |
| Signer B | 8.70 | 0 | 57.14 |

The result extracted from the proposed algorithm, as illustrated in both the Fig. 7 and Fig. 8, proves the system efficiency of above 90% that is competitive with recent research works [4-8].

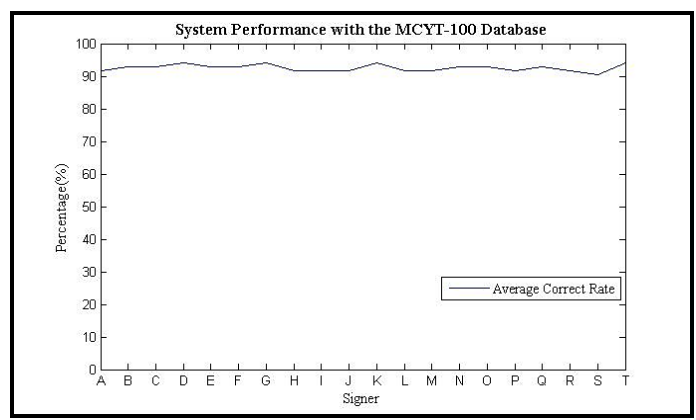

Fig. 7. System performance based on MCYT-100 database.

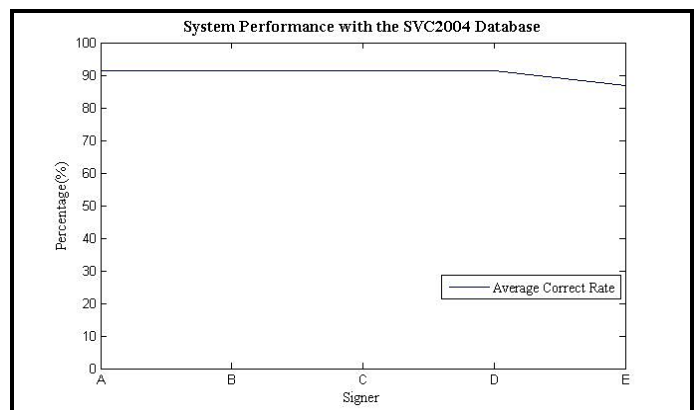

Fig. 8. System performance based on SVC2004 database.

## V. CONCLUSION

A signature verification system for DSP application was designed and implemented. The database of SVC2004 and MCYT-100 with cooperation of DWT and DCT provide an efficient structure to verify the authentic signatory. The parameters such as X-Y vector scaling, time and pressure of signing were considered in PCA platform to train the proposed algorithm. In addition, MATLAB interfacing was designed for user friendly. The analytical result taken from proposed algorithm based on both the SVC2004 and MCYT-100 database proved the system efficiency of 91.3% resolution in comparison with previous research works [4-8]. Further statistical experiments have been performed and the result of 8.7% in equal error rate is competent in DSP applications.